\documentclass[sigconf, nonacm, screen]{acmart}

\newcommand{\TOOLNAME}[1]{\textbf{\textit{RecruitScope}}}

\newcommand{\VIEW}[1]{\textbf{#1}}

\newcommand{\INDUSTRY}[1]{\textsc{Ind}-\textit{#1}}
\newcommand{\POSITION}[1]{\textsc{Pos}-\textit{#1}}
\newcommand{\EDUCATION}[1]{\textit{#1}}
\newcommand{\EXPERIENCE}[1]{\textit{#1}}
\newcommand{\PROVINCE}[1]{\textsc{Prov}-\textit{#1}}
\newcommand{\CITY}[1]{\textsc{City}-\textit{#1}}

\newcommand{\FIGURE}[1]{Fig.~\ref{#1}}

\newcommand{\USER}[1]{\textit{P}\textit{#1}}

\AtBeginDocument{%
  }

\acmConference[ICHEC 2025]{International Conference on Human-Engaged Computing}{November 21--23, 2025}{Singapore}

\acmSubmissionID{170}

\begin{document}

\title{\TOOLNAME{}: A Visual Analytics System for Multidimensional Recruitment Data Analysis}


\author{Xiyuan Zhu}
\orcid{0009-0007-2059-2110}
\affiliation{%
  \institution{Huazhong University of Science and Technology}
  \city{Wuhan}
  \state{Hubei}
  \country{China}
}
\email{AddInistrator.FallinDepart@gmail.com}

\author{Wenhan Lyu}
\orcid{0009-0007-6335-5166}
\affiliation{%
  \institution{Huazhong University of Science and Technology}
  \city{Wuhan}
  \state{Hubei}
  \country{China}
}
\email{3980181523@qq.com}

\author{Chaochao Fu}
\orcid{0009-0000-8684-0155}
\affiliation{%
  \institution{Huazhong University of Science and Technology}
  \city{Wuhan}
  \state{Hubei}
  \country{China}
}
\email{2460468415@qq.com}

\author{Yilin Wang}
\orcid{0009-0004-1622-9691}
\affiliation{%
  \institution{Huazhong University of Science and Technology}
  \city{Wuhan}
  \state{Hubei}
  \country{China}
}
\email{2208469290@qq.com}

\author{Jie Zheng}
\orcid{0009-0004-3498-8931}
\affiliation{%
  \institution{Huazhong University of Science and Technology}
  \city{Wuhan}
  \state{Hubei}
  \country{China}
}
\email{2639840745@qq.com}

\author{Qiyue Tan}
\orcid{0009-0000-3168-9734}
\affiliation{%
  \institution{Huazhong University of Science and Technology}
  \city{Wuhan}
  \state{Hubei}
  \country{China}
}
\email{2033935380@qq.com}

\author{Qianhe Chen}
\orcid{0000-0001-8291-5789}
\affiliation{%
  \institution{Huazhong University of Science and Technology}
  \city{Wuhan}
  \state{Hubei}
  \country{China}
}
\email{qianhechen01@gmail.com}

\author{Yixin Yu}
\orcid{0009-0003-5193-8592}
\affiliation{%
  \institution{Huazhong University of Science and Technology}
  \city{Wuhan}
  \state{Hubei}
  \country{China}
}
\email{sakaaanayu@gmail.com}

\author{Ran Wang}
\orcid{0000-0002-4340-0018}
\affiliation{%
  \institution{Huazhong University of Science and Technology}
  \city{Wuhan}
  \state{Hubei}
  \country{China}
}
\affiliation{%
  \department{Philosophy and Social Sciences Laboratory of Big Data and National Communication Strategy, Ministry of Education}
  \city{Wuhan}
  \state{Hubei}
  \country{China}
}
\email{rex_wang@hust.edu.cn}

\authornote{Corresponding author.}

\authorsaddresses{}


\begin{abstract}
  Online recruitment platforms have become the dominant channel for modern hiring, yet most platforms offer only basic filtering capabilities, such as job title, keyword, and salary range. This hinders comprehensive analysis of multi-attribute relationships and job market patterns across different scales. We present \TOOLNAME{}, a visual analytics system designed to support multidimensional and cross-level exploration of recruitment data for job seekers and employers, particularly HR specialists. Through coordinated visualizations, \TOOLNAME{} enables users to analyze job positions and salary patterns from multiple perspectives, interpret industry dynamics at the macro level, and identify emerging positions at the micro level. We demonstrate the effectiveness of \TOOLNAME{} through case studies that reveal regional salary distribution patterns, characterize industry growth trajectories, and discover high-demand emerging roles in the job market.
\end{abstract}





\maketitle

\begin{figure*}
  \centering
  \includegraphics[width=\textwidth]
  {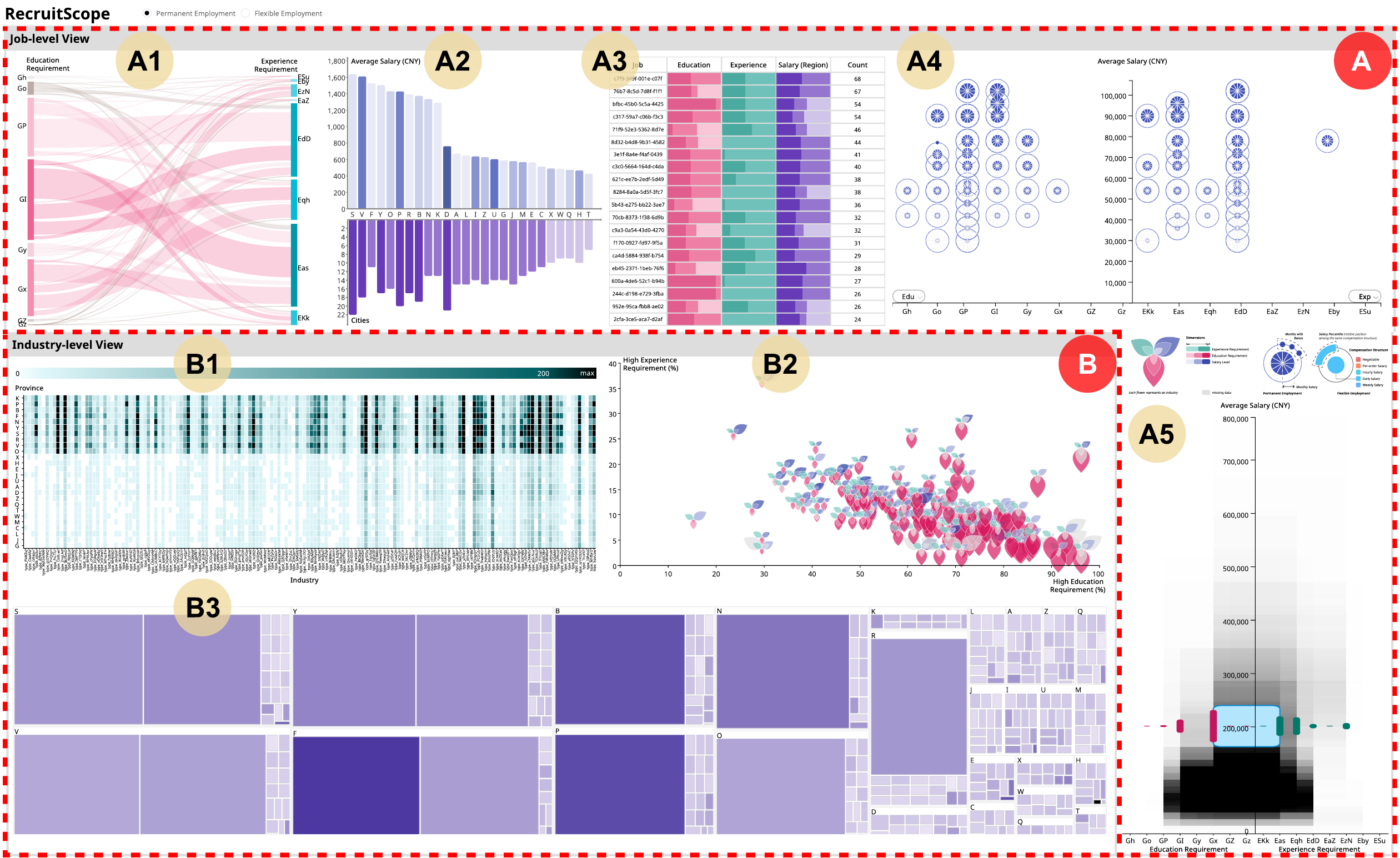}
  \caption{\TOOLNAME{}:, a multi-level visual analytics system for recruitment data. A) \VIEW{Job-level View} integrates education, experience, salary, and regional distribution to support detailed exploration of positions. B) \VIEW{Industry-level View} enables cross-industry and cross-regional comparison to uncover industry dynamics and salary patterns. Job titles, education and experience requirements, and regions shown in the figure are anonymized for data privacy.}
  \Description{teaser}
  \label{fig:teaser}
\end{figure*}

\section{Introduction}

Online recruitment platforms have become the dominant channel for modern hiring, where employers and job seekers increasingly rely on these platforms to post and search for job positions \cite{DillahuntIsraniLuCaiHsiao-ExaminingTheUseOfOnlinePlatformsForEmploymentASurveyOfUSJobSeekers-2021}. 
The widespread adoption of these platforms has generated a large amount of structured recruitment data with multiple dimensions such as job roles, salaries, locations, and skill requirements. 
Effective visualization of this multidimensional information can provide comprehensive insights into the job market.

However, existing visualization methods for recruitment data exhibit notable limitations. 
From the platform functionality perspective, most online job markets fail to adequately integrate job seekers' multifaceted requirements, including salary, location, and benefits packages, to name but a few. 
As a result, they cannot provide comprehensive insights into position-related information \cite{ChenPan-ResearchOnDataAnalysisAndVisualizationOfRecruitmentPositionsBasedOnTextMining-2022, WangChenWangXiongShen-JobvizSkill-drivenVisualExplorationOfJobAdvertisements-2024}. A fundamental reason for the limitations above is that current systems typically allow users to examine at most two attributes simultaneously and reduce interdependencies among multiple factors to binary relationships \cite{YokoyamaOkadaTaniguchi-PanaceaVisualExplorationSystemForAnalyzingTrendsInAnnualRecruitmentUsingTime-varyingGraphs-2021}. 
Consequently, conventional approaches struggle to reveal intricate interactions and latent patterns across multiple dimensions of recruitment records.

These limitations hinder an overall comprehension of the job market. 
On the one hand, it is difficult for users to simultaneously examine interactions across multiple dimensions, which limits their in-depth understanding of job differences and salary distribution patterns. 
On the other hand, conventional analytic methods make it difficult to infer industry dynamics and evolving trends in the job market. 
More importantly, when switching perspectives between the overall industry landscape and the details of specific positions, existing analytical tools lack sufficient flexibility and thus prevent users from developing a holistic view of the recruitment market.

To address the aforementioned limitations, we propose:

\begin{itemize}
    \item \TOOLNAME{}, a visual analytics system for recruitment data that supports exploration across multiple dimensions. It helps users develop a layered understanding of job market dynamics, from macro-level industry trends to micro-level job differences, thereby overcoming the limitations in analyzing interactions among multiple factors.
    \item \textbf{A novel flower-shaped scatterplot design}, inspired by the growing trend of glyph-based visualization, where each data entry is encoded as a compact visual mark to represent multiple attributes. In recruitment data analysis, it is crucial to demonstrate both the overall relationships among different industries or regions and the multidimensional composition of requirements within each category. Therefore, we integrate the spatial distribution of scatterplots with the multidimensional encoding capability of radar charts. Each petal encodes industry-specific requirements such as education, experience, and salary, while the spatial placement on the scatterplot reflects the relative position of industries within the recruitment market. This design not only captures the relative relationships among industries but also reveals internal requirement patterns within each industry, addressing the limitations of traditional radar charts that only support the comparison of a few objects.
    \item \textbf{Archetypal case studies} to validate the effectiveness of our system by revealing salary distribution patterns, identifying industry evolution trends, and discovering emerging job roles.
\end{itemize}

\section{Related Work}

\subsection{Labor Market Analytics}

\textbf{Labor Market Analytics (LMA)} leverages data, metrics, and statistical models to provide insights, forecasts, and decision support on labor supply and demand. It helps guide national employment policies \cite{GiabelliMalandriMercorioMezzanzanica-GraphlmiADataDrivenSystemForExploringLaborMarketInformationThroughGraphDatabases-2022} and assists enterprises in making targeted talent acquisition decisions \cite{QinZhangChengZhaShenZhangChenSunZhuZhuXiong-AComprehensiveSurveyOfArtificialIntelligenceTechniquesForTalentAnalytics-2025}.

The analytical foundations of LMA originated from traditional economic models, such as the Diamond-Mortensen-Pissarides model based on worker heterogeneity \cite{BhattacharyaJacksonCJenkins-RevisitingUnemploymentInIntermediateMacroeconomicsANewApproachForTeachingDiamond-mortensen-pissarides-2018}. With the development of the Internet, online recruitment data has become an important source for LMA \cite{MytnaKurekovaBeblavýThum-Thysen-UsingOnlineVacanciesAndWebSurveysToAnalyseTheLabourMarketAMethodologicalInquiry-2015}, offering up-to-date indicators of job market trends. However, conventional approaches often rely on manually labeled datasets with fixed patterns, which may fail to capture the dynamics of the labor market \cite{MurthyVelagaSrikaraPrabhasSattiKarthikBalivadaAttaBattula-AutomaticHrRecruitmentSystemAMultistageEvaluationApproach-2024}. Later, new approaches have been applied in LMA, such as dynamic graphs, hypernetworks, and synthetic augmentation \cite{ChenQinWangChengWangZhuXiong-Pre-dygaePre-trainingEnhancedDynamicGraphAutoencoderForOccupationalSkillDemandForecasting-2024, GuoLiuZhangZhangZhuXiong-TalentDemand-supplyJointPredictionWithDynamicHeterogeneousGraphEnhancedMeta-learning-2022, WolfNeubürgerLanwehr-GeneratingSyntheticDataForBetterPredictionModelingInSkillDemandForecasting-2023}.

Nevertheless, existing studies still face notable limitations. On the one hand, many studies remain focused on macro-level statistics, such as job counts or frequency analyze of specific skills, which fail to uncover complex interdependencies across multiple dimensions \cite{FaboKurekova-MethodologicalIssuesRelatedToTheUseOfOnlineLabourMarketData-2022}. On the other hand, labor market analysis requires identifying macro trends at the industry level while also capturing micro-level details at the level of individual job positions \cite{QinZhangChengZhaShenZhangChenSunZhuZhuXiong-AComprehensiveSurveyOfArtificialIntelligenceTechniquesForTalentAnalytics-2025}. Yet, current methods often concentrate on a single scale, making it difficult to support cross-level exploration \cite{FaboKurekova-MethodologicalIssuesRelatedToTheUseOfOnlineLabourMarketData-2022, ChenQinFangWangZhuZhuangZhuXiong-Job-sdfAMulti-granularityDatasetForJobSkillDemandForecastingAndBenchmarking-2024}. 

To address these limitations, we propose \TOOLNAME{}, a visual analytics system for recruitment big data that enables users to conduct multidimensional, cross-level exploration.

\subsection{Job Post Visualization}

\textbf{Job Post Visualization} refers to interactive visual analytics systems built on online recruitment data \cite{GutierrezCharleerDeCroonHtunGoetschalckxVerbert-ExplainingAndExploringJobRecommendationsAUser-drivenApproachForInteractingWithKnowledge-basedJobRecommenderSystems-2019, CharleerGutierrezVerbert-SupportingJobMediatorAndJobSeekerThroughAnActionableDashboard-2019}. It transforms large volumes of job postings into multidimensional structured information and presents them visually to assist both job seekers and employers \cite{MujtabaMahapatra-MiningAndAnalyzingOccupationalCharacteristicsFromJobPostings-2020}.

Existing work in visualization design has evolved from static displays to interactive analysis. Early prototypes were often static or offered limited interactivity, primarily focusing on keyword extraction and frequency analysis to identify basic relationships \cite{KamaruddinAbdulRahmanAmirah-Jobseeker-industryMatchingSystemUsingAutomatedKeywordSelectionAndVisualizationApproach-2019, SiLvYuanXiePeng-AnEfficientInterpretableVisualizationMethodOfMultidimensionalStructuralDataMatchingBasedOnJobSeekersAndPositions-2021}. In recent years, more big-data-driven, multi-view visual analytics prototypes have emerged, enabling linked exploration across multiple dimensions of recruitment data. 
Some studies further incorporate recommendation and matching explanations to enhance decision support and interpretability \cite{XinZhouLiu-PositionInformationVisualizationAnalysisAndPersonalizedRecommendationBasedOnAntColony-2024}. Meanwhile, systems such as JobViz \cite{WangChenWangXiongShen-JobvizSkill-drivenVisualExplorationOfJobAdvertisements-2024} reflect a shift from keyword search to skill-driven analysis in job exploration, while approaches such as Skills2Graph \cite{GiabelliMalandriMercorioMezzanzanicaSeveso-Skills2graphProcessingMillionJobAdsToFaceTheJobSkillMismatchProblem-2021} show the possibility of leveraging large-scale corpora to support multidimensional relationship exploration.

In conclusion, recruitment data contain rich information such as job position heterogeneity, salary patterns, and industry trends, but current analytical methods still struggle to reveal the intricate interrelations among these factors. Therefore, \TOOLNAME{} aims to uncover latent relationships within complex recruitment data, thereby providing better support for user decision-making in the job market.

\section{Design Goals}

Based on preliminary research findings, RecruitScope targets two core user groups within the recruitment ecosystem: job seekers and employers, particularly HR specialists. The system is designed to support job seekers in efficiently exploring career opportunities through multi-level visual analysis, while also providing employers and HR specialists with insights into the structure and trends of the talent market. We thus propose the following design goals:

From the perspective of job seekers, we propose \textbf{G1}, which is to \textbf{support multi-dimensional analysis of job position heterogeneity and salary patterns.}

\begin{itemize}
    \item \textbf{G1.1: Quantifying differences across multiple dimensions.} The system should integrate diverse features of recruitment records to help users evaluate differences among job positions. This enables a macro-level understanding of the demand structure in the recruitment market and provides a foundation for more informed decision-making and planning.
    \item \textbf{G1.2: Revealing latent salary distribution patterns.} Salaries in recruitment data often exhibit strong skewness, long-tail effects, or structural disparities, making it difficult to detect clusters of high-paying positions or areas of low-salary risk. The system should therefore visualize hidden salary distribution patterns to assist users in formulating effective policies and job-selection strategies.
\end{itemize}

From the perspective of employers or HR specialists, we propose \textbf{G2}, which is \textbf{to support hierarchical analysis from macro to micro levels to uncover latent insights from recruitment data.}

\begin{itemize}
    \item \textbf{G2.1: Macro-level explanation of industry dynamics.} Industry development typically follows cyclical patterns of growth, maturity, and decline \cite{OSullivan-IndustrialLifeCycleRelevanceOfNationalMarketsInTheDevelopmentOfNewIndustriesForEnergyTechnologies--TheCaseOfWindEnergy-2020}. However, existing analyze of recruitment data often focus only on short-term changes in posting counts, lacking a depiction of long-term trajectories. This makes it difficult for users to identify which industries are rising and which are in decline. The system should therefore reveal the evolutionary pathways of expanding and contracting industries from a macro-level perspective.
    \item \textbf{G2.2: Micro-level discovery of emerging positions with urgent talent demand.} Emerging positions are usually small in scale and widely dispersed, making them difficult to detect through conventional aggregate statistics. Nonetheless, they hold strong predictive value for labor market shifts. The system should therefore support micro-level, position-granular analysis to uncover rapidly growing job positions.
\end{itemize}

\section{Data Processing}

\begin{table*}[t]
  \caption{Final Data Structure}
  \label{tab:dataset-schema}
  \small
  \begin{tabular}{p{2cm} p{4cm} p{2cm} p{8cm}}
    \toprule
    Field & Definition & Number & Notation Example / Enumeration \\
    \midrule
    \_id & Unique code for job position & 1,713 &
      \POSITION{\texttt{([a-z0-9]{4}-){3}[a-z0-9]{4}}} \\

    province & Provincial-level administrative division & 26 &
      \PROVINCE{\texttt{[A-Z]}} \\

    city & Municipal-level administrative division & 370 &
      \CITY{\texttt{[A-Z][0-9]{3}}} \\

    salary & -- & -- & -- \\

    salary\_type & Compensation structure & 7 & -- \\

    salary\_base & Salary period for permanent employment & -- & -- \\

    upper\_bound & Upper bound of salary & -- & -- \\

    lower\_bound & Lower bound of salary & -- & -- \\

    company & Unique code for company & 136,149 & -- \\

    industry & Unique code for industry category & 158 &
      \INDUSTRY{\texttt{[A-Za-z0-9]{6}}} \\

    education & Educational requirements & 8 &
      \begin{tabular}[t]{@{}l@{}}
      \EDUCATION{Doctor (Gh)}, \EDUCATION{Master (Go)}, \EDUCATION{Bachelor (GP)}, \\
      \EDUCATION{Junior College (GI)}, \EDUCATION{High School (Gy)}, 
      \EDUCATION{Technical (Gx)}, \\
      \EDUCATION{Middle School (GZ)}, \EDUCATION{No requirement (Gz)}
      \end{tabular} \\

    experience & Work experience requirements & 8 &
      \begin{tabular}[t]{@{}l@{}}
      \EXPERIENCE{10+ yrs (ESu)}, \EXPERIENCE{8–9 yrs (Eby)}, 
      \EXPERIENCE{6–7 yrs (EzN)}, \EXPERIENCE{5 yrs (EaZ)}, \\
      \EXPERIENCE{3–4 yrs (EdD)}, \EXPERIENCE{2 yrs (Eqh)}, 
      \EXPERIENCE{1 yr (Eas)}, \EXPERIENCE{No requirement (EKk)}
      \end{tabular} \\
    \bottomrule
  \end{tabular}
  \normalsize
\end{table*}

We use recruitment data provided by 2024 ChinaVis Challenge\footnote{\href{https://chinavis.org/2024/challenge.html}{2024 ChinaVis Challenge}}, comprising 430,664 recruitment records that cover 169,540 job positions, 267,296 companies, and 158 industries, distributed across 26 provincial-level and 371 municipal-level divisions. The dataset contains 8 types of experience requirements and 8 types of education requirements. All data were anonymized; positions, companies, and industries were replaced with unique identifiers, while administrative divisions were represented using a scheme of one-letter provincial code with a three-digit municipal code. Personal information and other sensitive fields were removed. Throughout this paper, we use abbreviated notations to refer to these anonymized entities. Table~\ref{tab:dataset-schema} presents the complete notation system with representative examples.

\textbf{Data Filtering.} We first ranked positions by the number of associated recruitment records. The results show that 148,201 positions were linked to only one recruitment record, indicating a clear long-tail distribution. Since a single or very few records may reflect temporary or incidental hiring needs rather than stable market demand, we selected the top 1\% of positions by recruitment volume as the basis for analysis.

\textbf{Requirement Encoding.} We mapped the education requirements to three levels: high education (\EDUCATION{Bachelor's Degree(GP)} and above), medium education (\EDUCATION{Senior High School Diploma(Gy)} and \EDUCATION{Technical Secondary School Diploma(Gx)}), and low education (\EDUCATION{Junior High School Diploma(GZ)} and \EDUCATION{No Educational Requirements(Gz)}) \cite{NationalPeoplesCongressofthePeoplesRepublicofChina-HigherEducationLawOfThePeoplesRepublicOfChina-2021}. Likewise, experience requirements were merged into two levels: high experience ($\geq$5 years) and low experience ($<$5 years).

\textbf{Salary Normalization.} Salaries in the original data appeared in various formats (annual, monthly, weekly, etc.). We first categorized employment type into permanent employment and flexible employment, and then extracted the upper and lower salary bounds \cite{PeoplesRepublicofChina-LaborContractLawOfThePeoplesRepublicOfChina-2008}. All salaries were converted into an annual basis \cite{HumanResourcesPeoplesRepublicofChinaMOHRSS-NoticeOfTheMinistryOfHumanResourcesAndSocialSecurityOnIssuesConcerningTheCalculationOfEmployeesAverageMonthlyWorkingHoursAndWagesThroughoutTheYear-2025}.

\textbf{Outlier Detection.} Given the structural differences between permanent and flexible employment, outlier removal was performed separately for the two groups. Recruitment records were first grouped by provincial-level division and job position, and the Interquartile Range (IQR) method was applied within each group.
\begin{align}
L &= Q_1 - 1.5 \times \mathrm{IQR} \\
U &= Q_3 + 1.5 \times \mathrm{IQR}
\end{align}
where $Q1$ and $Q3$ denote the 25th and 75th percentiles, respectively. Records outside the interval $[L, U]$ were considered outliers and removed. We adopted the IQR method because it does not assume normality and has been widely used in recruitment and salary analysis due to its robustness \cite{ChenXiao-SpreadRegressionSkewnessRegressionAndKurtosisRegressionWithAnApplicationToTheUsWageStructure-2025}.

The record structure of the final dataset is shown in Table~\ref{tab:dataset-schema}.

\begin{figure*}[ht]
    \centering
    \includegraphics[width=\linewidth]{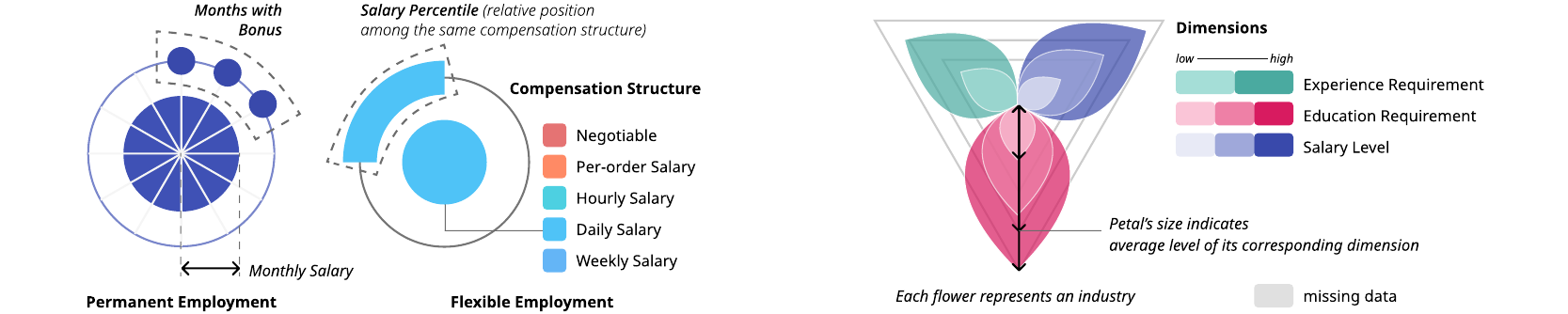}
    \caption{Glyph design in (left) \VIEW{Salary Pattern Scatterplot} and (right) \VIEW{Industry Requirements Distribution View}}
    \Description{Petal-shaped glyph for industry representation.}
    \label{fig:glyph}
\end{figure*}

\section{System Overview}

Based on the design goals, we developed \TOOLNAME{}, a visual analytics system consisting of eight coordinated views, organized into two levels of analysis: the \VIEW{Job-level View (A)} and the \VIEW{Industry-level View}

\subsection{\VIEW{Job-level View}}

The \VIEW{Job-level View} helps users quantify differences between job positions across multiple dimensions \textbf{(G1.1)} and quickly grasp overall hiring patterns \textbf{(G1.2)}. It contains five parts:

\VIEW{Education--Experience Sankey Diagram (A1)}. 
Education requirements and work experience requirements are the two most frequently occurring categories in job postings, usually presented together to set candidates’ entry thresholds. The \VIEW{Education–Experience Sankey Diagram} visualizes the correspondence between education and experience requirements across job positions. Rectangles on the left represent education requirements, while those on the right denote experience requirements. The width of the flows between them indicates the number of recruitment records matching each pair of education and experience conditions. Specific pairs of education and experience requirements can be filtered by selecting the corresponding rectangle labels or flows. Compared with alternatives such as heatmaps or grouped bar charts, the Sankey diagram reveals the matching patterns between these two categorical dimensions more effectively\cite{VosoughKammerKeckGroh-MirroringSankeyDiagramsForVisualComparisonTasks-2018} and offers better visual continuity, which facilitates understanding of the overall distribution of entry thresholds for positions.

\VIEW{Region--City Bidirectional Bar Chart (A2)}.
The \VIEW{Region--City Bidirectional Bar Chart} presents the recruitment scale and regional structure across provincial-level administrative divisions. The horizontal axis represents provincial divisions. The upper bar chart encodes the number of recruitment records in each province, with opacity representing the average salary level of the corresponding province. The lower bar chart represents the number of municipal-level divisions within each province.

\VIEW{Job Comparison List (A3)}.
Education and experience requirements primarily define entry thresholds for positions, while regional recruitment distribution patterns reveal market trends and salary disparities. Compared with examining the education–experience relationship \textbf{(A1)} or regional/cross-industry salary distribution \textbf{(A2)} separately, the \VIEW{Job Comparison List} integrates education, experience, and regional salary distributions into a unified view. This extended grouped bar chart design incorporates multiple attribute dimensions within a compact layout, enabling efficient comparison of combined features. The view is linked with the \VIEW{Education--Experience Sankey Diagram} and the \VIEW{Region--City Bidirectional Bar Chart}, allowing users to focus on filtered data entries. The area of each rectangle encodes the proportion of each category within the selected job position, opacity encodes requirement levels in the ``Education'' and ``Experience'' columns, and distinguishes among high, medium, and low salary tiers in the ``Salary(Region)'' column.

\VIEW{Salary Pattern Scatterplot (A4)}.
The \VIEW{Salary Pattern Scatterplot} addresses the need to support visual comparison between two types of compensation structures \textbf{(G1.2)}: permanent employment and flexible employment. In recruitment data, the former is typically characterized by monthly salaries, while the latter includes various forms such as weekly, daily, and hourly wages. To meet this requirement, the scatterplot preserves the strengths of conventional scatterplots in revealing patterns and outliers, while introducing a ring-based encoding to integrate multiple dimensions of information within limited visual space. The horizontal axis allows users to select one dimension for its positive and negative sides via a dropdown menu. Each point corresponds to a single recruitment data entry. The glyph design is shown in \FIGURE{fig:glyph}. For permanent employment, the size of the inner circle encodes the monthly salary, while the number of small circles on the outer ring indicates the number of bonus months provided. For flexible employment, color distinguishes among weekly, daily, and hourly wages, while the arc on the outer ring shows the relative position of a record’s salary within all recruitment records of the same type.

\VIEW{Job Requirements Distribution View (A5)}.
The \VIEW{Job Requirements Distribution View} extends the box plot to better represent the relationship between discrete and continuous attributes in recruitment data. Since education requirements and experience requirements are categorical while salary is continuous, the whiskers used in standard box plots are not suitable. To address this, the design replaces lines with bands, enabling a more intuitive depiction of education, experience, and salary's distribution across job positions. Each colored block represents a single position: its horizontal span corresponds to the ranges of education and experience associated with that position, and its vertical span encodes the corresponding salary range. Overlapping blocks are visually aggregated, with higher opacity indicating denser overlaps. Each band along the horizontal axis corresponds to one education or experience level, with its length representing the proportion of positions requiring that level. By integrating entry requirements and market rewards within a single view, this design enhances the comparability of multidimensional recruitment data and facilitates quick identification of concentration areas, disparities, and outlier trends in salary distributions.

\subsection{Industry-level View}

The \VIEW{Industry-level View} emphasizes macro-level comparisons across industries and regions. It supports the analysis of industry dynamics from multiple perspectives, including salary patterns, job demand, and requirement distributions \textbf{(G2.1)}, and it facilitates the discovery of emerging positions with urgent talent demand \textbf{(G2.2)}.

\VIEW{Industry--Region Distribution View (B1)}.
The \VIEW{Industry--Region Distribution View} ranks provincial-level administrative divisions and industries based on salary levels, revealing geographic preferences of different industries and industry preferences of different regions. The x-axis ranks industries by average salary in descending order, while the y-axis ranks the 26 provincial-level units similarly. The color opacity in each grid cell represents the number of recruitment data records.

\VIEW{Industry Requirements Distribution View (B2)}.
The \VIEW{Industry Requirements Distribution View} aims chart to analyze industries across three dimensions: education requirements, experience requirements, and salary. Each flower-shaped glyph represents an industry: red petals encode education, green petals encode experience, blue petals encode salary, and gray petals indicate missing values. Petal size reflects the average level of the corresponding attribute. The x- and y-axes represent the proportions of recruitment records within each industry that require high education and high experience, respectively.
We extend the conventional radar chart for visual design. Radar charts are a classic method for visualizing multidimensional data but face significant challenges in large-scale industry comparisons. When hundreds of industries need to be displayed simultaneously, overlapping polygons severely reduce readability. 
Radar charts also emphasize overall contour comparisons among a small number of objects, making it difficult to quickly identify differences in specific dimensions. Grouping industries and arranging multiple radar charts in a grid partially alleviates overlap, but each chart becomes too small to discern details, and the grid layout fragments industries into isolated units, obscuring their relative positions in the broader market. 
Alternative designs have been proposed, such as \textbf{Origami Plots} \cite{DuanTongSuttonAschChuSchmidChen-OrigamiPlotANovelMultivariateDataVisualizationToolThatImprovesRadarChart-2023} and \textbf{Flow Radar Glyphs} \cite{HlawatschLeubeNowakWeiskopf-FlowRadarGlyphs—staticVisualizationOfUnsteadyFlowWithUncertainty-2011}. Origami Plots mitigate axis-order sensitivity by fixing coordinate axis sequences but still suffer from overlap in large-scale scenarios. Flow Radar Glyphs highlight temporal and directional features in flow-field data, but dense glyph placement often leads to severe visual clutter. These approaches are therefore ill-suited to large-scale heterogeneous recruitment data. 
To overcome these obstacles, we propose the \textbf{flower-shaped scatterplot design}. By encoding each industry as a compact flower glyph and leveraging spatial distribution in the scatterplot, the design avoids overlap and preserves visual clarity at scale. It enables users to rapidly detect anomalies in individual dimensions while maintaining an organized two-dimensional distribution of industries. As a result, the view simultaneously reveals the relative positioning of industries within the recruitment market as well as the internal composition of requirements for each industry.

\VIEW{Regional Profile View (B3)}.
The \VIEW{Regional Profile View} displays either a \textbf{provincial-level (B3.1)} or \textbf{municipal-level (B3.2)} division using a treemap layout to show the distribution of average salary, high-demand positions, and industry preferences. In the provincial-level regional view, each gray rectangle represents a provincial unit, containing smaller rectangles for its constituent city-level units. The area of each rectangle reflects the number of job postings, while its color opacity represents the average salary relative to other regions. The municipal-level view, accessed by clicking on a primary region, uses the same treemap scheme.Within each treemap cell, a three-layer donut chart is embedded: the outer and middle rings respectively show the top five job positions and top five industries by proportion, while the inner ring’s opacity encodes the average salary level. This hierarchical structure naturally aligned with the nested administrative divisions makes the treemap a desirable choice for visualizing multi-level regional distributions. The treemap efficiently utilizes space and conveys hierarchical relationships, while the embedded ring charts further overlay industry and position information within the regional framework, adding multidimensional context without compromising readability.\cite{SoaresMirandaLimaResquedosSantosMeiguins-DepictingMoreInformation-2020}
By visualizing regional, occupational, and industrial information together with salary levels, this view supports exploration from macro-level market structures to micro-level employment preferences, offering a comprehensive regional workforce profile.

\section{Archetypal Case Study}

We conducted two archetypal case studies to demonstrate the effectiveness of our proposed system.

\subsection{Case 1: Uncovering Hidden Opportunities in the Labor Market}

Job seeker \USER{1}, who holds a \EDUCATION{Bachelor’s Degree (GP)} and has \EXPERIENCE{3–4 years of work experience (EdD)}, hopes to secure a position that offers both high salary and promising career prospects. However, in the competitive recruitment market, he feels that his qualifications may put him at a disadvantage.

\begin{figure}[h]
    \centering
    \includegraphics[width=0.6\linewidth]{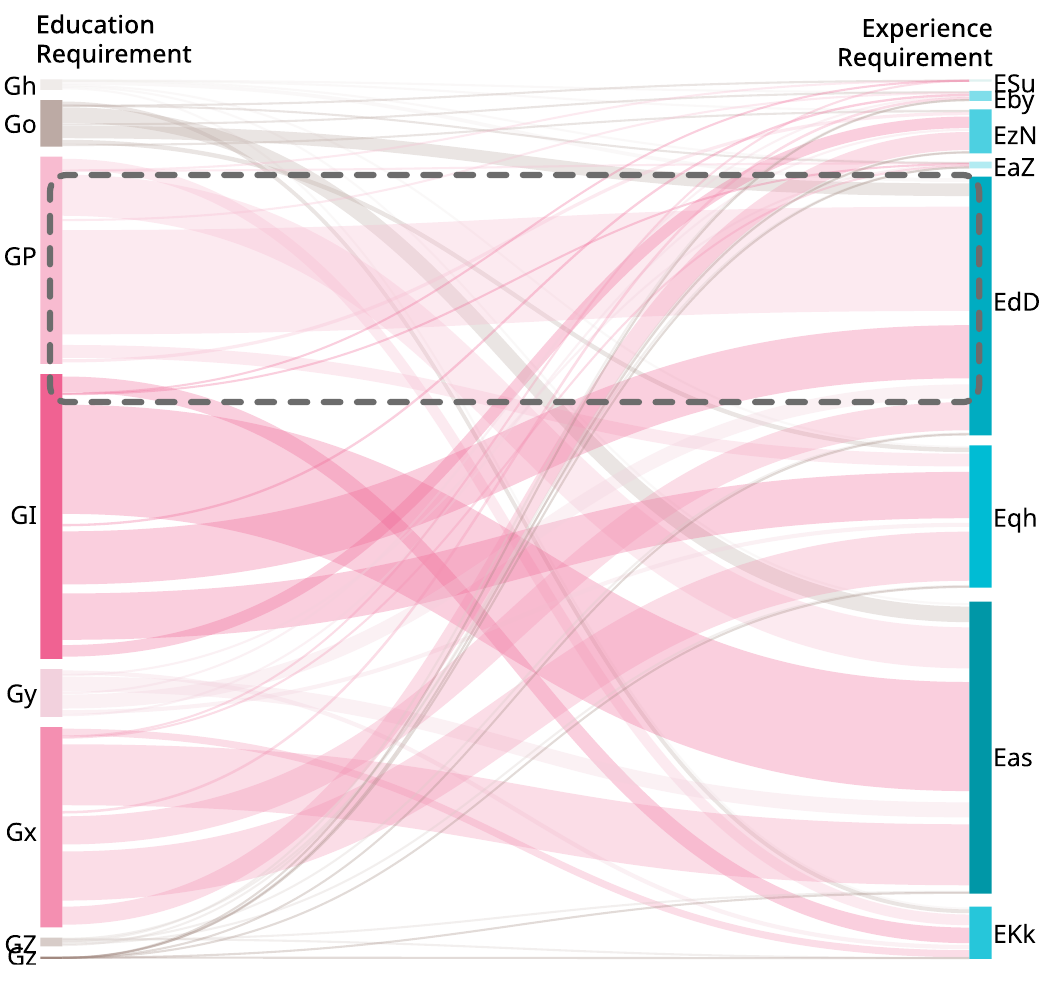}
    \caption{The flow \EDUCATION{GP}--\EXPERIENCE{EdD} is notably thicker than most other combinations, indicating substantial market demand for this qualification pairing.}
    \Description{The flow \EDUCATION{GP}--\EXPERIENCE{EdD} is notably thicker than most other combinations, indicating substantial market demand for this qualification pairing.}
    \label{fig:Case 1-1}
\end{figure}

\begin{figure}[htbp]
    \centering
    \includegraphics[width=0.6\linewidth]{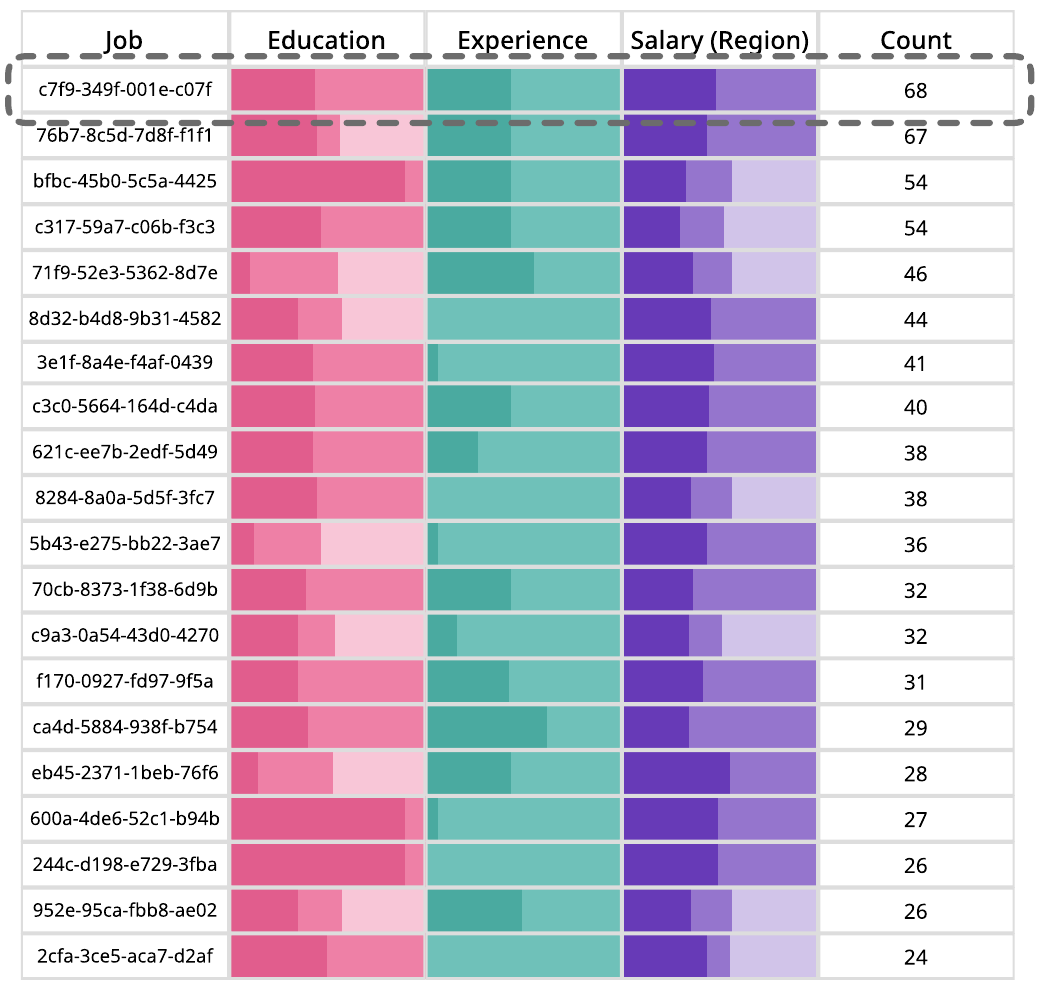}
    \caption{Job positions ranked by posting count for \EDUCATION{GP}--\EXPERIENCE{EdD}. The highlighted position (\POSITION{c7f9-349f-001e-c07f}) has the highest frequency with 68 postings.}
    \Description{Job positions ranked by posting count for \EDUCATION{GP}--\EXPERIENCE{EdD}. The highlighted position (\POSITION{c7f9-349f-001e-c07f}) has the highest frequency with 68 postings.}
    \label{fig:Case 1-2}
\end{figure}

\begin{figure*}[htbp]
    \centering
    \includegraphics[width=0.6\linewidth]{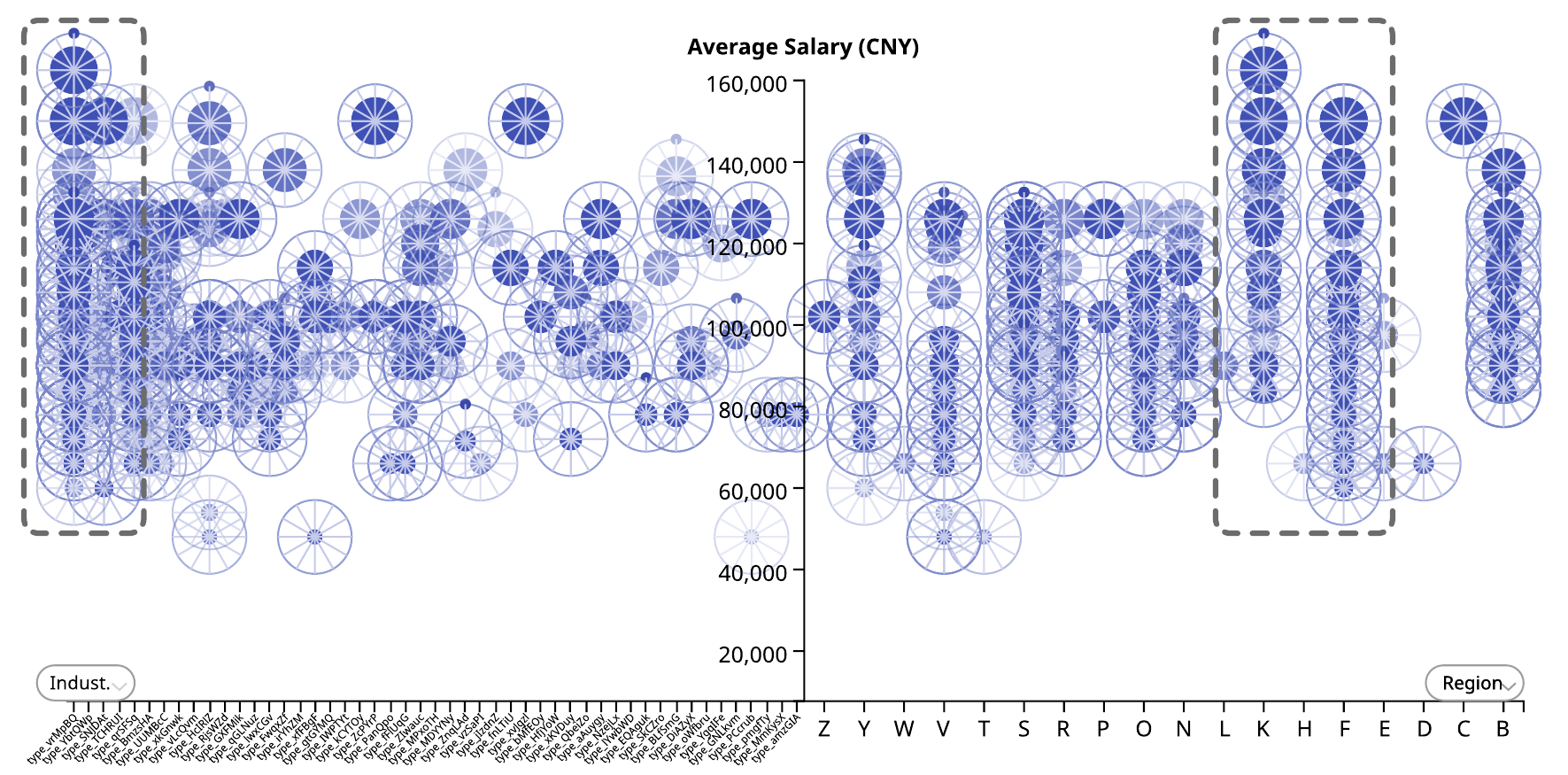}
    \caption{The selected position (\POSITION{c7f9-349f-001e-c07f}) concentrates in \INDUSTRY{vrMpBQ} and \INDUSTRY{YbtQwp} with higher salaries in \PROVINCE{K} and \PROVINCE{F}.}
    \Description{The selected position (\POSITION{c7f9-349f-001e-c07f}) concentrates in \INDUSTRY{vrMpBQ} and \INDUSTRY{YbtQwp} with higher salaries in \PROVINCE{K} and \PROVINCE{F}.}
    \label{fig:Case 1-3}
\end{figure*}

\begin{figure*}[ht]
    \centering
    \includegraphics[width=0.6\linewidth]{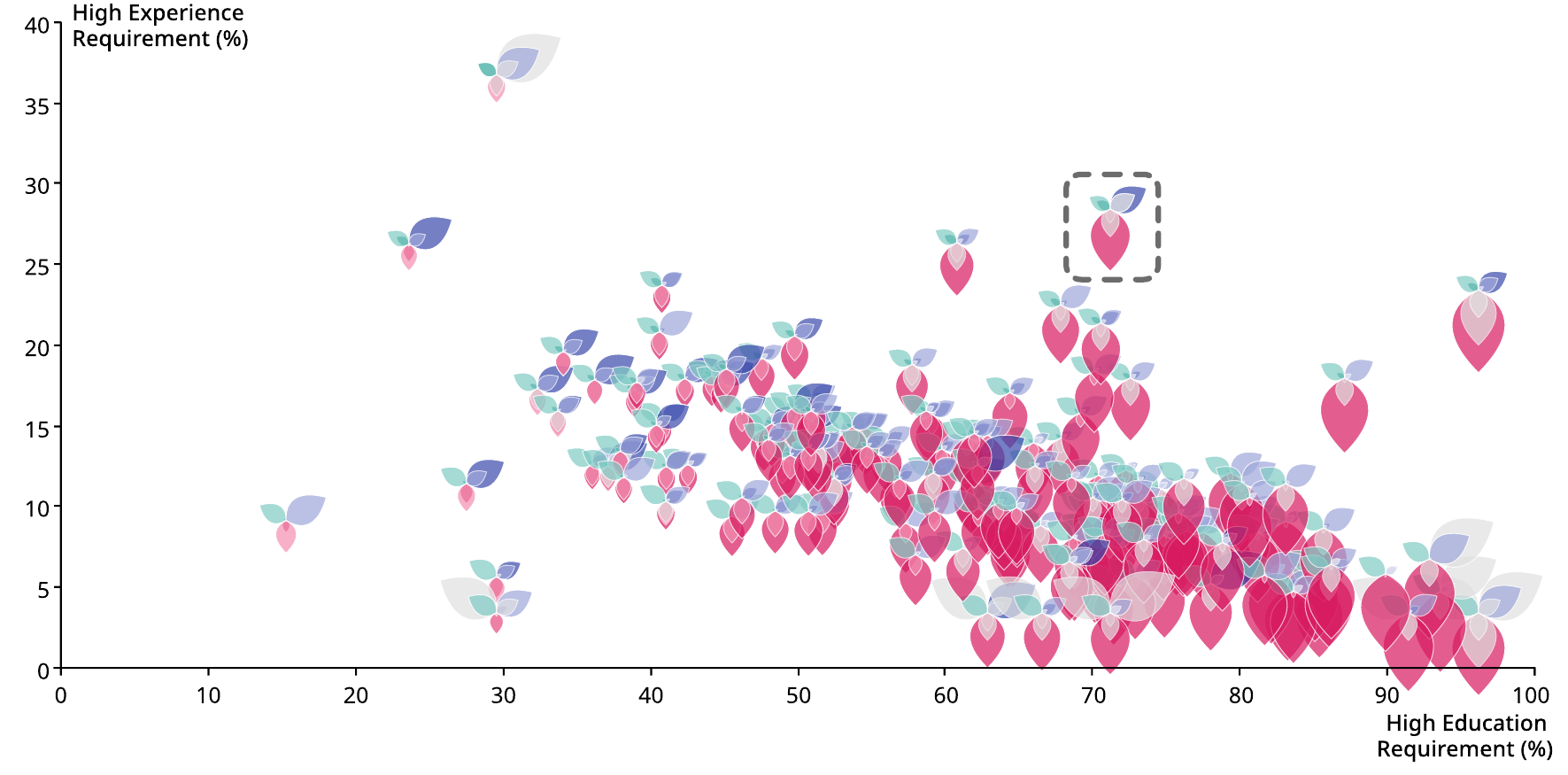}
    \caption{\INDUSTRY{qrSFSq} appears in the upper-right area with large pink petals, suggesting high qualification requirements and strong demand for permanent positions.}
    \Description{\INDUSTRY{qrSFSq} appears in the upper-right area with large pink petals, suggesting high qualification requirements and strong demand for permanent positions.}
    \label{fig:Case 1-4}
\end{figure*}

\USER{1} turns to the \VIEW{Education--Experience Sankey Diagram} and selects the combination of \EDUCATION{GP} and \EXPERIENCE{EdD}. The relatively thick flow indicates a substantial market demand, which slightly eases his concerns and motivates him to further explore the distribution of positions and industries.

\textbf{Industry Differences and Opportunities.} After selecting \EDUCATION{GP}--\EXPERIENCE{EdD} in \VIEW{Education--Experience Sankey Diagram}, the system displays all relevant positions in the \VIEW{Job Comparison List}. \USER{1} selects the top-ranked position, \POSITION{c7f9-349f-001e-c07f}, for deeper analysis. In the \VIEW{Salary Pattern Scatterplot}, this position is concentrated mainly in the \INDUSTRY{vrMpBQ} and \INDUSTRY{YbtQwp} industries, with relatively high salary levels in \PROVINCE{K} and \PROVINCE{F}, ranging from 50,000 to 160,000 CNY annually. Most recruitment records offer fixed monthly salaries, while some provide an additional one-month bonus.

To further assess career prospects, \USER{1} switches to the \VIEW{Salary Pattern Scatterplot} and observes that the \INDUSTRY{qrSFSq} industry is densely concentrated in economically developed regions, signaling that it is an expanding industry. In the \VIEW{Industry–Region Distribution View}, the scatter point of \INDUSTRY{qrSFSq} appears in the middle-right area with a relatively large pink petal, indicating that the industry is entering a growth phase.

\begin{figure}
    \centering
    \includegraphics[width=0.6\linewidth]{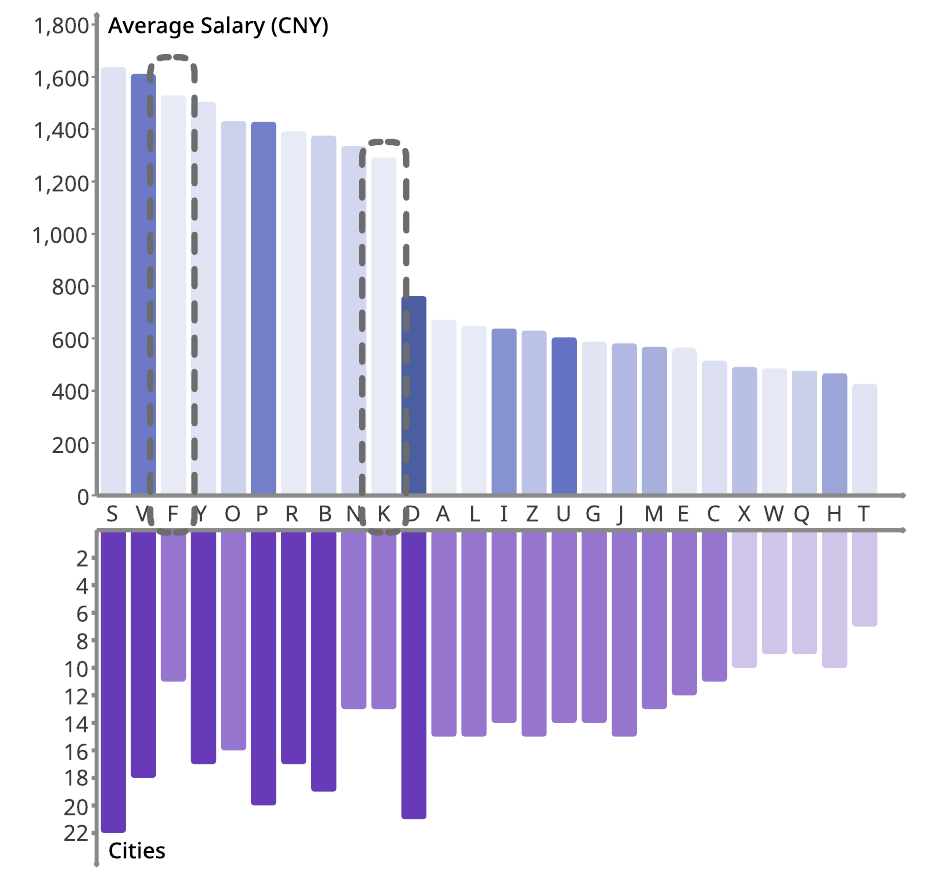}
    \caption{\PROVINCE{K} and \PROVINCE{F} demonstrate higher average salaries.}
    \label{fig:Case 1-5}
    \Description{\PROVINCE{K} and \PROVINCE{F} demonstrate higher average salaries.}
\end{figure}

\begin{figure*}
    \centering
    \includegraphics[width=0.8\linewidth]{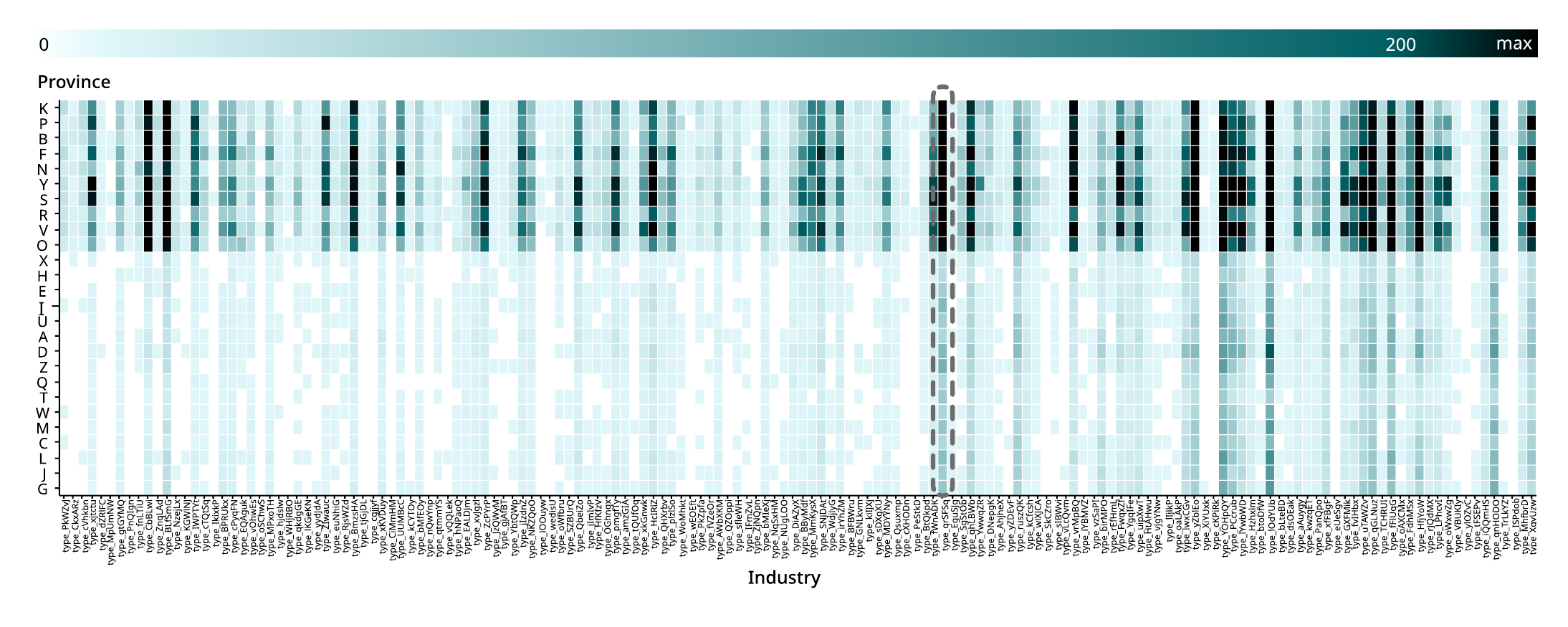}
    \caption{\INDUSTRY{qrSFSq} clusters in darker cells representing economically developed regions with substantial job availability.}
    \Description{\INDUSTRY{qrSFSq} clusters in darker cells representing economically developed regions with substantial job availability.}
    \label{fig:Case 1-6}
\end{figure*}

\begin{figure*}
    \centering
    \includegraphics[width=\linewidth]{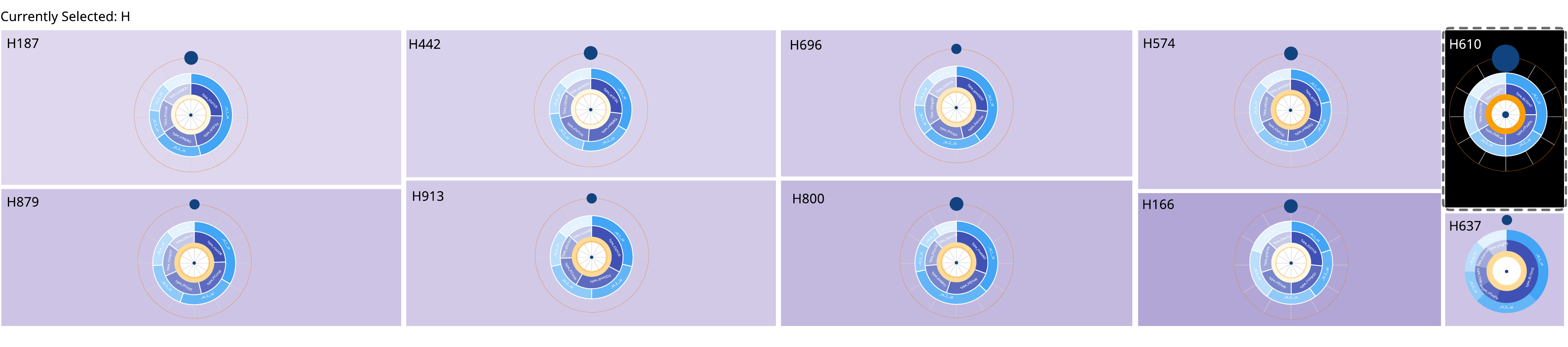}
    \caption{\CITY{H610} stands out with remarkably higher salary intensity than neighboring cities.}
    \Description{\CITY{H610} stands out with remarkably higher salary intensity than neighboring cities.}
    \label{fig:Case 2-1}
\end{figure*}

\textbf{The Importance of Regional Choice}. \USER{1} initially planned to target top-tier cities. However, the \VIEW{Region--City Bidirectional Bar Chart} shows that \PROVINCE{K} and \PROVINCE{F} offer substantially higher average salaries than those in other regions. Combining this with insights from the \VIEW{Industry--Region Distribution View}, \USER{1} confirms that salaries for the selected position rank among the highest in \PROVINCE{K} and \PROVINCE{F}. This reshapes his career-choosing approach, highlighting that geographic choice directly impacts salary.

\textbf{Reflection and Decision-Making}. \USER{1} realizes that traditional job-seeking strategies often overemphasize education and experience while overlooking other critical dimensions, such as the stage of industry development, regional salary disparities, and compensation structures. Taking these factors into account, \USER{1} decides to focus on opportunities in \INDUSTRY{qrSFSq} within \PROVINCE{K} and \PROVINCE{F}, where salaries are comparatively high and industry prospects are strong. Moreover, since most job positions offer a fixed monthly salary with limited bonuses, \USER{1} plans to emphasize base salary during salary negotiations.

\begin{figure}
      \centering
      \includegraphics[width=0.6\linewidth]{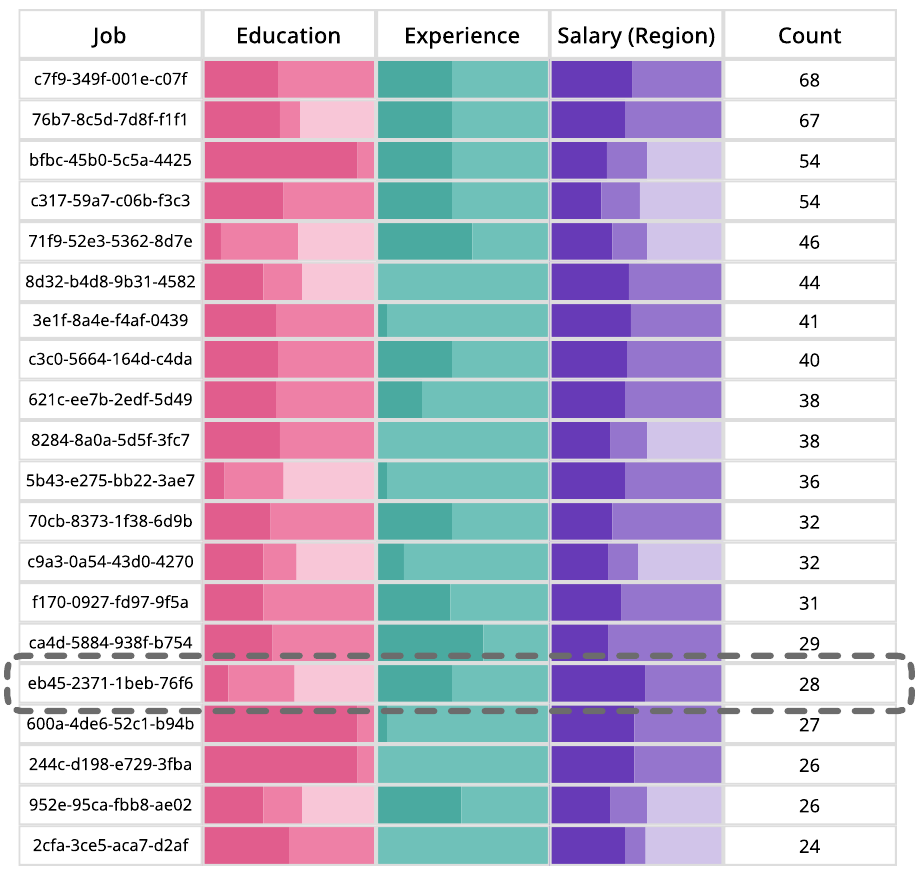}
      \caption{\POSITION{eb45-2371-1beb-76f6} ranks among the highest salaries but requiring only medium education and low experience levels.}
      \label{fig:Case 2-2}
      \Description{\POSITION{eb45-2371-1beb-76f6} ranks among the highest salaries but requiring only medium education and low experience levels.}
\end{figure}

\subsection{Case 2: Detecting Talent Gap Signals in Emerging Positions}

Technology company \USER{2} in \PROVINCE{H} seeks to identify niche labor markets with relatively low competition in order to to design differentiated recruitment strategies and gain an advantage in the competitive talent landscape. \USER{2} assumes that job positions offering abnormally high salaries but requiring relatively low entry thresholds often signal a shortage of labor.

\textbf{Salary Anomaly in \CITY{H610}.} Examining the overall recruitment landscape of \PROVINCE{H} through the \VIEW{Regional Profile View}, \USER{2} observes that although the province’s total recruitment volume is not the highest, one municipal-level division stands out with significantly higher salary levels than those in surrounding areas. This anomaly captures the company’s attention, and \USER{2} reasons that such elevated salaries indicate a labor supply shortage for specific positions.

\begin{figure*}
      \centering
      \includegraphics[width=0.6\linewidth]{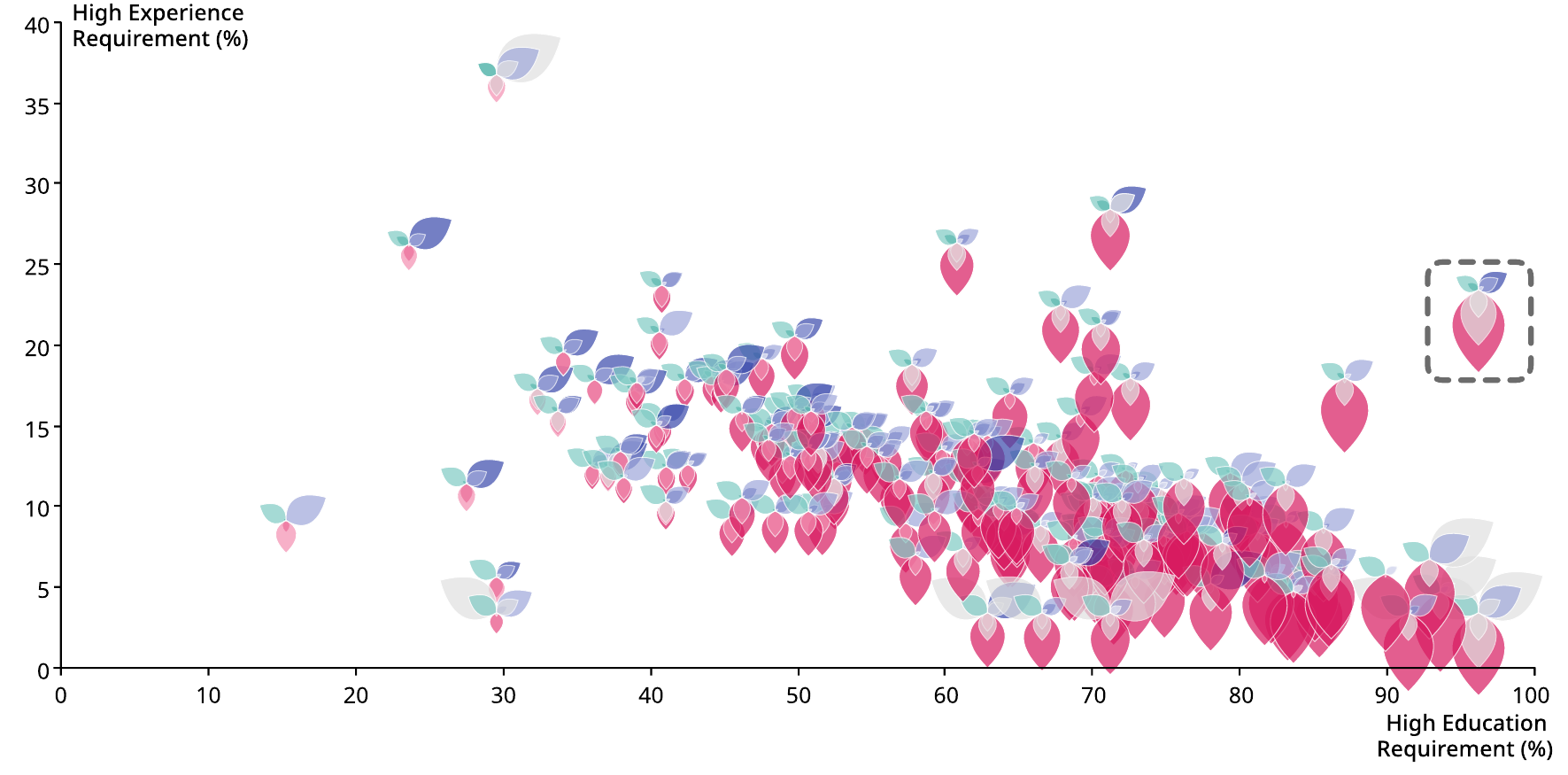}
      \caption{The industry's marker appears in the upper-right quadrant with a large pink petal, indicating it is a fast-growing, technology-intensive industry.}
      \label{fig:Case 2-4}
      \Description{The industry's marker appears in the upper-right quadrant with a large pink petal, indicating it is a fast-growing, technology-intensive industry.}
\end{figure*}

\begin{figure}
    \centering
    \includegraphics[width=0.6\linewidth]{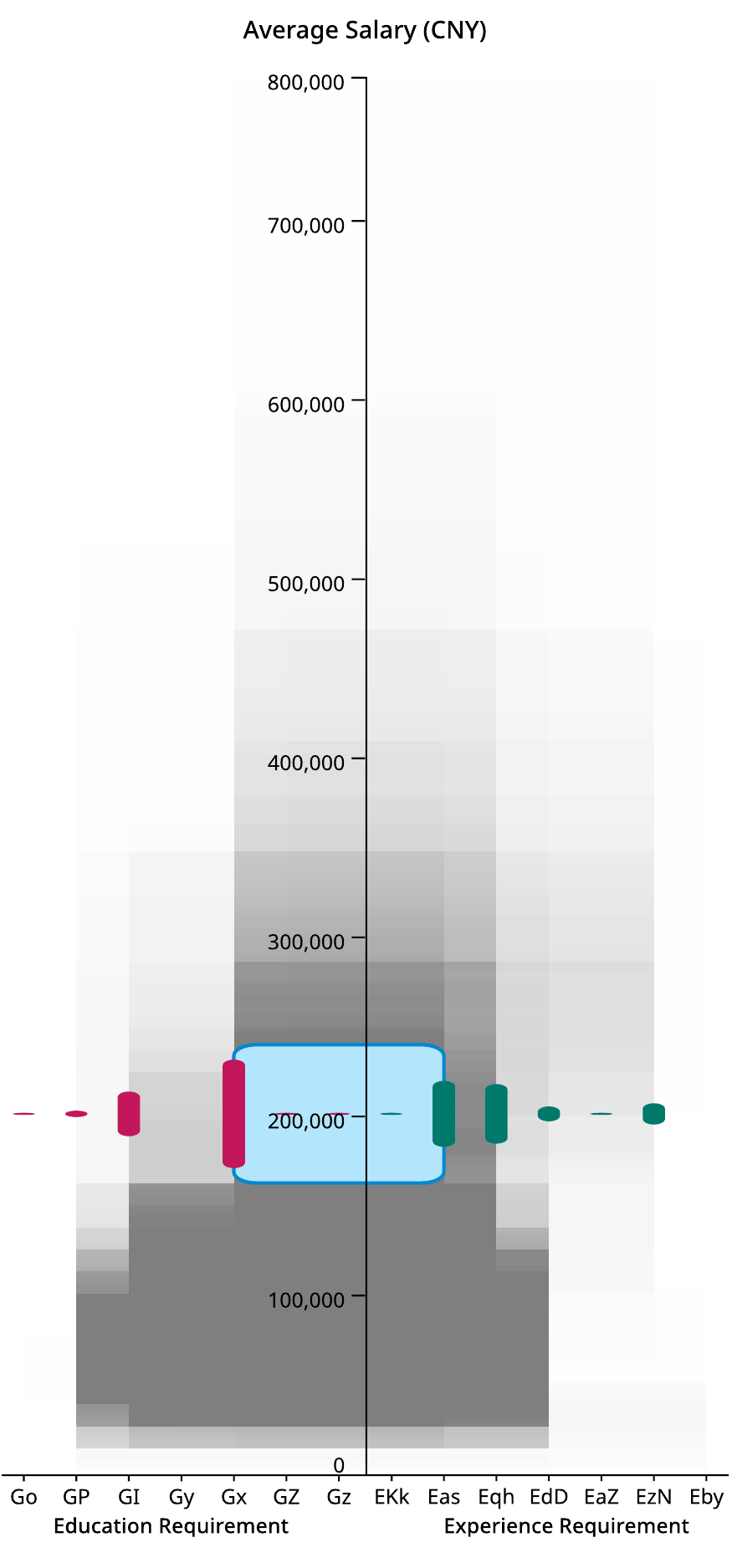}
    \caption{\POSITION{eb45-2371-1beb-76f6} occupies the medium-education, low-experience quadrant, representing a high-salary opportunity with moderate entry barriers.}
    \label{fig:Case 2-3}
    \Description{\POSITION{eb45-2371-1beb-76f6} occupies the medium-education, low-experience quadrant, representing a high-salary opportunity with moderate entry barriers.}
\end{figure}

\textbf{Salary Compression.} \USER{2} then selects the pair of \EDUCATION{Technical Secondary School Diploma (Gx)} and \EXPERIENCE{1 year of work experience (Eas)} in the \VIEW{Education–Experience Sankey Diagram}. Combining insights from the \VIEW{Job Comparison List} and the \VIEW{Job Requirements Distribution View}, \USER{2} finds that \POSITION{eb45-2371-1beb-76f6} ranks among the highest in salary, yet its requirements fall into the medium-education and low-experience categories. Further analysis in the \VIEW{Salary Pattern Scatterplot} reveals that the annual salary for this position ranges from 80,000 to 150,000 CNY, significantly higher than that of other positions with comparable education and experience requirements.

\textbf{Emerging Job Positions.} \USER{2} subsequently analyzes the industry of this position in the \VIEW{Industry Requirements Distribution View} and finds that the industry’s scatterpoint appeared in the upper-right area, with a large pink petal, indicating a fast-growing, technology-intensive industry. This suggests that the position is not only currently scarce but may also become an emerging position in the long-term development of the industry.

\textbf{Reflection and Decision-Making.} Based on the analysis, the enterprise formulates the following recruitment strategies. First, \USER{2} decides to lower the education requirements for \POSITION{eb45-2371-1beb-76f6}. Second, \USER{2} plans to establish partnerships with vocational and technical schools to secure outstanding graduates from relevant programs in advance. Finally, \USER{2} intends to provide a competitive starting salary complemented with a skill training system to attract and retain talent.

\section{Conclusion and Future Work}

We propose \TOOLNAME{}, a visual analytics system for recruitment data that supports analysis from industry down to individual jobs, which can help reveal multidimensional patterns and trends in the recruitment market. On the one hand, \TOOLNAME{} combines data analysis and visual presentation into a single platform. On the other hand, our study addresses the challenge of maintaining readability of multidimensional data at scale through a novel flower-shaped scatterplot design. It reveals complex relationships across multiple dimensions, including job position heterogeneity, compensation structures, and geographical characteristics, thereby providing new perspectives for understanding labor market dynamics.

However, \TOOLNAME{} still has some limitations. First, although the system can be applied in real-world scenarios, its ability to handle recruitment datasets with different structures and to generalize across sources still needs to be validated. Second, while \TOOLNAME{} supports coordinated exploration across multiple attributes, it lacks compact summarization. For example, nonlinear relationships among education level, experience, and salary (e.g. higher education does not always correspond to higher pay) often require cross-view reasoning to discover.

In future work, we will extend the system to more diverse and realistic datasets, integrating heterogeneous data from multiple platforms and incorporating richer fields such as job descriptions, required skills, and benefit packages. We will also investigate improved visual encodings and interaction techniques to lower the learning curve and increase the system’s overall usability.

\begin{acks}
This work is supported in part by the National Social Science Foundation of China, under Grant 25BXW041; in part by the Fundamental Research Funds for the Central Universities, Huazhong University of Science and Technology, under Grant 82400049.
\end{acks}


\bibliographystyle{ACM-Reference-Format}
\bibliography{reference}


\end{document}